\documentclass{article}
\usepackage{arxiv}

\usepackage[utf8]{inputenc} % allow utf-8 input
\usepackage[T1]{fontenc}    % use 8-bit T1 fonts
\usepackage{amsmath}      % blackboard math symbols
\usepackage[hyphens]{url}        % simple URL typesetting
\usepackage{graphicx}
\RequirePackage{filecontents}    
\usepackage{booktabs}       % professional-quality tables
\usepackage{amsfonts} 
\usepackage{booktabs}
\usepackage{nicefrac}       % compact symbols for 1/2, etc.
\usepackage{microtype}      % microtypography
\usepackage{lipsum}
\usepackage{fancyhdr}       % header
\usepackage[table]{xcolor}
\usepackage{authblk}
\usepackage{doi}
\usepackage[numbers]{natbib}
\usepackage{hyperref}
\usepackage{cleveref}

\title{Distributed Transfer Learning with 4\textsuperscript{th} Gen Intel\textsuperscript{\textregistered} Xeon\textsuperscript{\textregistered} Scalable Processors}

\author{\thanks{{\{lakshmi.arunachalam, fahim.mohammad, vrushabh.h.sanghavi\}}@intel.com}
	{Lakshmi Arunachalam, Fahim Mohammad, Vrushabh H. Sanghavi}}

\affil {AI Framework Engineer\\ Data Center and Artificial Intelligence \\ 
	Intel Corporation} 

\renewcommand{\shorttitle}{Distributed Transfer Learning using Intel\textsuperscript{\textregistered} Xeon\textsuperscript{\textregistered} Processors.}

\hypersetup{
	pdftitle={Distributed Transfer Learning with 4\textsuperscript{th} Gen Intel\textsuperscript{\textregistered} Xeon\textsuperscript{\textregistered} Scalable Processors},
	pdfsubject={cs.AI, cs.CV, cs.CL },
	pdfauthor={Lakshmi Arunachalam, Fahim Mohammad, Vrushabh Sanghavi},
		pdfkeywords={Artificial Intelligence, Deep Learning Optimization, End-to-End AI applications, E2E performance optimization, Transfer Learning, Intel\textsuperscript{\textregistered} Xeon\textsuperscript{\textregistered}}
	}

\begin{document}
	
	\maketitle
	
	\begin{abstract}
		In this paper, we explore how transfer learning, coupled with Intel\textsuperscript{\textregistered} Xeon\textsuperscript{\textregistered}, specifically 4th Gen Intel\textsuperscript{\textregistered} Xeon\textsuperscript{\textregistered} scalable processor, defies the conventional belief that training is primarily GPU-dependent. We present a case study where we achieved near state-of-the-art accuracy for image classification on a publicly available Image Classification TensorFlow dataset using Intel\textsuperscript{\textregistered} Advanced Matrix Extensions(AMX) and distributed training with Horovod.
	\end{abstract}

% keywords can be removed

\section{Introduction}

	Imagine how kids learn to start coloring with crayons. It may take few days for them to learn how to hold the crayon, stay within the picture and so on. They may need lots of crayons and coloring books so that they can get the hang of it. Then they can easily apply their skills to learn how to use color pencils, painting, pencil shading or master art. They don't have to start all over again from scratch because they already have a foundation of coloring with crayons. This is what transfer learning is about. Instead of starting from scratch and needing more time and resources, we can use the skills already learnt and finetune a little more to learn a similar task and be good at it. 
	
	In the world of machine learning and artificial intelligence, transfer learning has emerged as a powerful technique.  In this blog, we explore how transfer learning, coupled with Intel\textsuperscript{\textregistered} Xeon\textsuperscript{\textregistered} Scalable CPUs, specifically 4th Gen Intel\textsuperscript{\textregistered} Xeon\textsuperscript{\textregistered} scalable processor, defies the conventional belief that training is primarily GPU-dependent. We present a case study where we achieved near state-of-the-art accuracy for image classification on a publicly available Image Classification TensorFlow dataset \cite{tfdataset} using Intel\textsuperscript{\textregistered} Advanced Matrix Extensions(AMX)\cite{intelamx} and distributed training with Horovod.

\section{Image Classification with Transfer Learning}
	\label{sec:e2eaiapp}
\begin{figure}
				\centering
				\includegraphics[width=\textwidth]{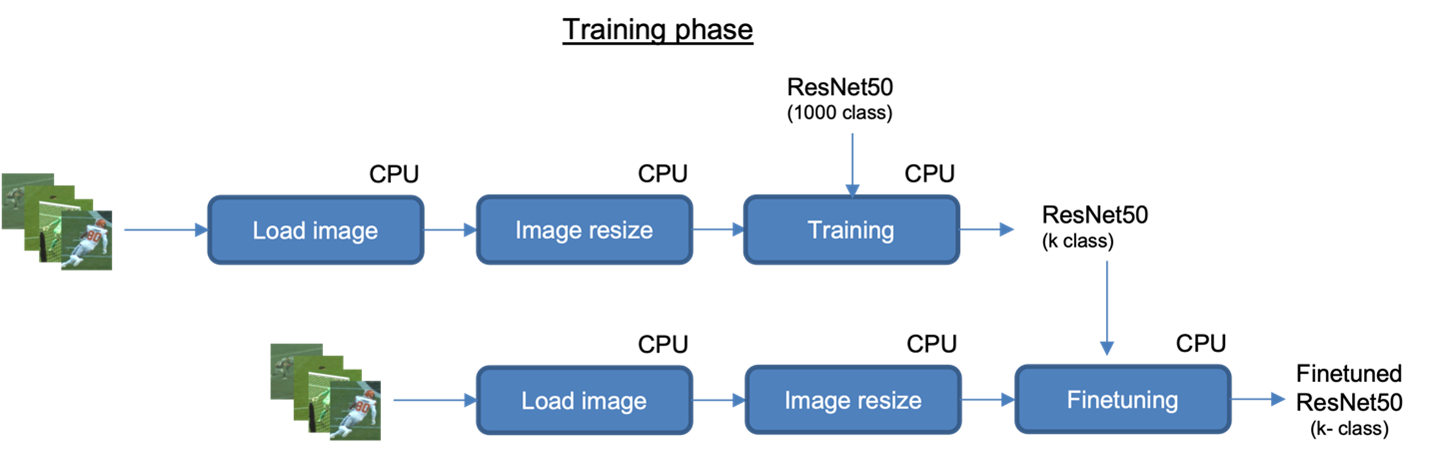}
				\caption{Vision Transfer Learning Pipeline}
				\label{fig:fig1}
			\end{figure}

	The basic idea behind transfer learning is to use a pre-trained model, often trained on a large and diverse dataset, as a starting point. This pre-trained model has already learned useful features or representations from its original task, which can be transferred and applied to the new task. The advantage of transfer learning lies in its ability to significantly reduce the time and resources needed for training while delivering impressive results. To illustrate the power of transfer learning, let's consider a case study of identifying colorectal cancer tissue types through image classification \cite{kather2016multi}. We started with the pre-trained ResNet v1.5 \cite{resnet50v15-intel}\cite{resnet50-he2015} weights and fine-tuned the last classification layer using a TensorFlow dataset with 5000 images with 4000 for training. This approach allowed us to build on the knowledge acquired during pre-training and achieve close to state-of-the-art accuracy of 94.5\% \cite{10153365} on this dataset. Data augmentation was used as a preprocessing step, and early stopping criteria with a patience of 10 was employed to stop training once convergence was reached. The pipeline demonstrated run-to-run variations of 6-7 epochs, with an average of 45 epochs to achieve convergence. Figure ~\ref{fig:fig1} shows the transfer learning pipeline on vision task.

	\section{Leveraging Intel\textsuperscript{\textregistered} Xeon\textsuperscript{\textregistered} Scalable CPUs}
		Traditionally, training deep learning models was GPU-intensive, but with Intel\textsuperscript{\textregistered} Xeon\textsuperscript{\textregistered} Scalable CPUs, we witnessed a paradigm shift. By utilizing Intel\textsuperscript{\textregistered} Advanced Matrix Instructions (AMX) with BF16 precision, we achieved remarkable accuracy of 94.5\% with the model converging in just 43 epochs. The entire training process took less than 5.5 minutes with a single socket, showcasing the efficiency and speed of Intel\textsuperscript{\textregistered} Optimization for TensorFlow. Intel\textsuperscript{\textregistered} Optimization for TensorFlow is powered by Intel\textsuperscript{\textregistered} oneAPI Deep Neural Network Library (oneDNN) \cite{intel-onednn} \cite{intel-oneapi}, which includes convolution, normalization, activation, inner product, and other primitives vectorized. To achieve the above performance and accuracy, we used the following settings:

		\begin{itemize}
			\item Use Mixed Precision: Leverage Intel\textsuperscript{\textregistered} AMX BF16 precision format by enabling auto mixed precision in TensorFlow. BF16 offers better precision than FP16 while maintaining higher performance than FP32. In our case study we achieved similar accuracy with BF16 as with FP32.
			
			\item Use \texttt{numactl}: Accessing memory from the local socket is faster than from a remote socket in NUMA systems. To avoid potential performance issues due to remote memory access, bind the memory to one socket using the \texttt{numactl} command. For hyperthreading scenarios, use the command \texttt{numactl -C 0-55,112-167 -m 0 python train.py} to ensure memory is bound to one socket.
			
			\item Define run time parameters: Inter-op parallelism involves distributing tasks across cores to manage system resources efficiently and improve overall system performance. Intra-op parallelism focuses on optimizing parallel execution within a single core, breaking tasks into smaller sub-tasks to boost performance in single-threaded applications. For the case study, the inter-op parallelism is set to 56 threads (number of cores), and the intra-op parallelism is set to 56 threads. Additionally, use specific KMP settings as below
				\begin{itemize}
					\item  \texttt{KMP\_SETTINGS = 1}
					\item  \texttt{KMP\_BLOCKTIME = 1}
					\item  \texttt{OMP\_NUM\_THREADS = NUM\_CORES (56)}
					\item  \texttt{KMP\_AFFINITY = granularity=fine,compact,1,0}
				\end{itemize}
		\end{itemize}
		
	\section{Empowering Multi-Socket Performance with Distributed Training}
\begin{figure}
					\centering
					\includegraphics[width=\textwidth]{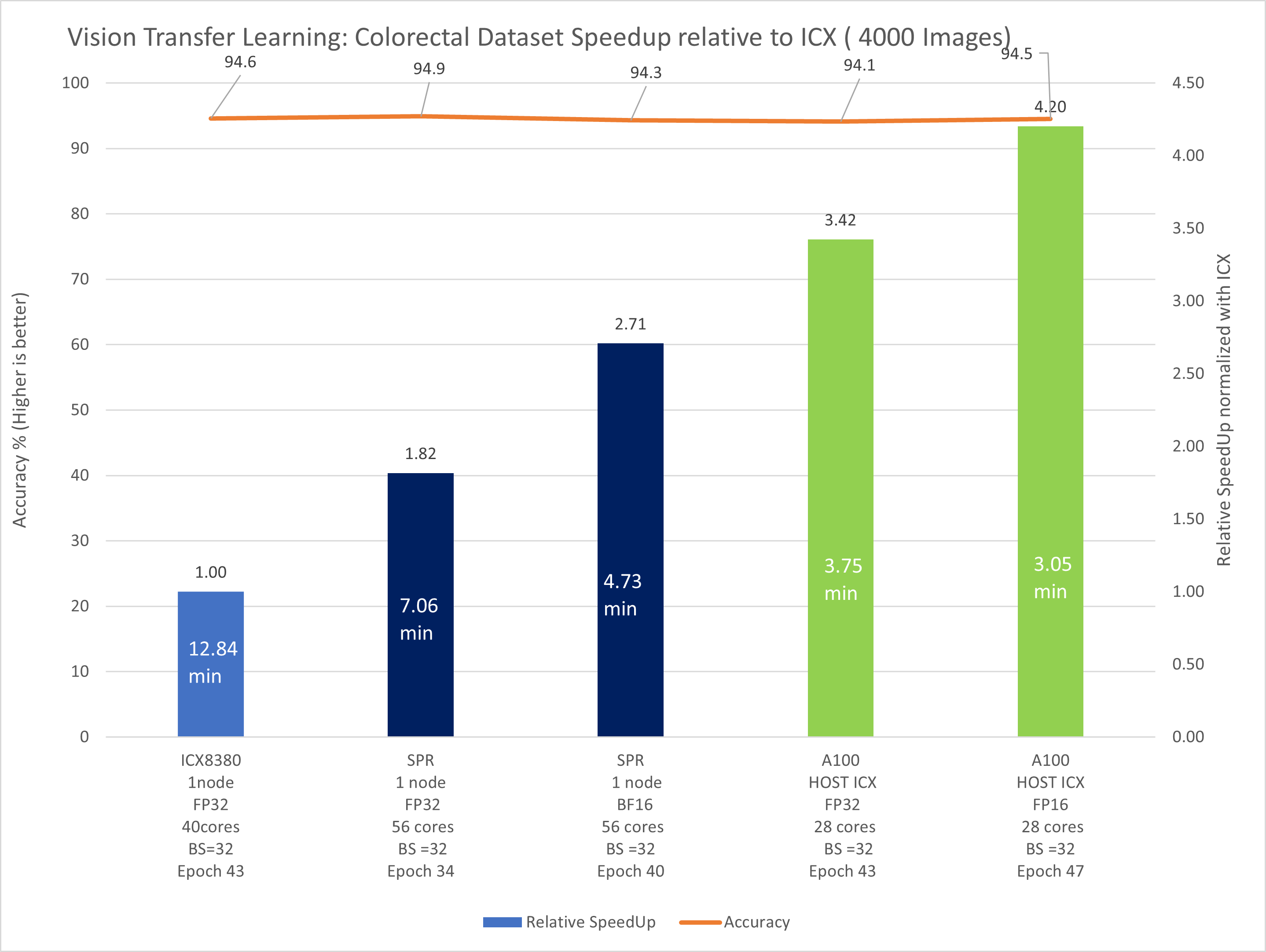}
					\caption{Competitive Perf Results for Vision Transfer Learning Workload.}
					\label{fig:fig2}
				\end{figure}

		Intel\textsuperscript{\textregistered} Xeon\textsuperscript{\textregistered} Scalable CPUs come equipped with two sockets, each having 56 cores. To maximize performance, we employed distributed training with Horovod \cite{sergeev2018horovod} and OpenMPI \cite{open-mpi} as the backend. Horovod, an open-source distributed training framework developed by Uber, supports popular deep learning frameworks like TensorFlow, PyTorch, and MXNet. By leveraging MPI, Horovod efficiently distributes training data and model parameters across multiple devices, resulting in faster training times. With all 112 cores, including hyperthreading, fully engaged, we achieved an impressive training time of around 3 minutes, comparable to an out-of-the-box training on an NVIDIA A100 Rome GPU. The total training time results are displayed in Figure ~\ref{fig:fig2}.
		
		In the specified distributed training setup, weak scaling is used, maintaining the same batch size throughout. The training is performed using Horovod with two workers on a single Sapphire Rapids system, where each worker is mapped to one socket of the system. The dataset is divided into halves and assigned to each worker for processing.
		
		To reduce communication overhead, gradients are averaged every 5 epochs instead of after each epoch. The training process utilizes the Horovod optimizer, and a warmup period of 3 epochs is set. The initial learning rate is set to 0.01, and it is scaled by the number of workers to 0.002. To optimize intra-op parallelism, the number of threads is set to 54, which is the number of cores minus 2. This configuration aims to achieve efficient and effective training while leveraging the computational capabilities of the Sapphire Rapids system.

		To maximize the performance on Intel\textsuperscript{\textregistered} Xeon\textsuperscript{\textregistered} CPU leverage we followed the recipe below.
	
\section{Conclusion}
Transfer learning has proven to be a game-changer in deep learning, enabling us to build on existing knowledge and achieve outstanding results with minimal time and resources. The successful application of transfer learning on Intel\textsuperscript{\textregistered} Xeon\textsuperscript{\textregistered} Scalable CPUs, particularly Sapphire Rapids, challenges the GPU-centric training mindset and offers a compelling alternative for high-performance image classification tasks. As we continue to explore the possibilities of leveraging Intel\textsuperscript{\textregistered}'s advanced technologies, we look forward to even greater strides in AI and machine learning.

\section*{Configuration Details}
	\begin{itemize}
		\item {3rd Gen Intel\textsuperscript{\textregistered} Xeon\textsuperscript{\textregistered} scalable processor (ICX)}
		Test by Intel\textsuperscript{\textregistered} as of 10/21/2022. 1-node, 2x Intel\textsuperscript{\textregistered} Xeon\textsuperscript{\textregistered} Platinum 8380, 40 cores, HT On, Turbo On, Total Memory 1024 GB (16 slots/ 64 GB/ 3200 MHz [run @ 3200 MHz] ), SE5C620.86B.01.01.0005.2202160810, 0xd000375, Ubuntu 22.04.1 LTS, 5.15.0-48-generic, n/a, Vision Transfer Learning Pipeline,Intel-tensorflow-avx512 2.10.0, resnet50v1\_5, n/a

	\item{4th Gen Intel\textsuperscript{\textregistered} Xeon scalable processor (SPR)}
		Test by Intel\textsuperscript{\textregistered} as of 10/21/2022. 1-node, 2x Intel\textsuperscript{\textregistered} Xeon\textsuperscript{\textregistered} Platinum 8480+ ,56 cores, HT On, Turbo On, Total Memory 1024 GB (16 slots/ 64 GB/ 4800 MHz [run @ 4800 MHz] ), EGSDREL1.SYS.8612.P03.2208120629 , 0x2b000041 , Ubuntu 22.04.1 LTS, 5.15.0-48-generic, n/a, Vision Transfer Learning Pipeline, Intel-tensorflow-avx512 2.10.0, resnet50v1\_5, n/a.

	\item{NVIDIA-A100}
		 Test by Intel\textsuperscript{\textregistered} as of 10/26/2022. 1-node (DGX-A100), 2xAMD EPYC 7742 64-Core Processor, 64 cores, HT On, Turbo On,, Total 1024GB (16 slots/64GB/3200 MHz) [run @ 3200MHz] ), Nvidia A100 GPU, BIOS 1.1, 0x830104d ,Ubuntu 20.04.2 LTS , 5.4.0-81-generic,  n/a, Vision Transfer Learning Pipeline, Tensorflow 2.10, resnet50v1\_5, n/a.
	\end{itemize}

%Bibliography
\bibliographystyle{unsrtnat}  
\bibliography{references}  

\begin{thebibliography}{10}
\providecommand{\natexlab}[1]{#1}
\providecommand{\url}[1]{\texttt{#1}}
\expandafter\ifx\csname urlstyle\endcsname\relax
  \providecommand{\doi}[1]{doi: #1}\else
  \providecommand{\doi}{doi: \begingroup \urlstyle{rm}\Url}\fi

\bibitem[tfd()]{tfdataset}
{TensorFlow Datasets: a collection of ready-to-use datasets.}
\newblock URL \url{https://www.tensorflow.org/datasets}.

\bibitem[int({\natexlab{a}})]{intelamx}
{Accelerate AI Workloads with Intel AMX}, {\natexlab{a}}.
\newblock URL
  \url{https://www.intel.com/content/www/us/en/products/docs/accelerator-engines/advanced-matrix-extensions/ai-solution-brief.html}.

\bibitem[Kather et~al.(2016)Kather, Weis, Bianconi, Melchers, Schad, Gaiser,
  Marx, and Z{"o}llner]{kather2016multi}
Jakob~Nikolas Kather, Cleo-Aron Weis, Francesco Bianconi, Susanne~M Melchers,
  Lothar~R Schad, Timo Gaiser, Alexander Marx, and Frank~Gerrit Z{"o}llner.
\newblock Multi-class texture analysis in colorectal cancer histology.
\newblock \emph{Scientific reports}, 6:\penalty0 27988, 2016.

\bibitem[res()]{resnet50v15-intel}
{Intel ResNet 50v1.5 models}.
\newblock URL
  \url{https://github.com/IntelAI/models/tree/master/benchmarks/image_recognition/tensorflow/resnet50v1_5}.

\bibitem[He et~al.(2015)He, Zhang, Ren, and Sun]{resnet50-he2015}
Kaiming He, Xiangyu Zhang, Shaoqing Ren, and Jian Sun.
\newblock {Deep Residual Learning for Image Recognition}, 2015.
\newblock URL \url{https://arxiv.org/abs/1512.03385}.

\bibitem[Plumworasawat and Sae-Bae(2023)]{10153365}
Sirithep Plumworasawat and Napa Sae-Bae.
\newblock Colorectal tissue image classification across datasets.
\newblock In \emph{2023 20th International Conference on Electrical
  Engineering/Electronics, Computer, Telecommunications and Information
  Technology (ECTI-CON)}, pages 1--4, 2023.
\newblock \doi{10.1109/ECTI-CON58255.2023.10153365}.

\bibitem[int({\natexlab{b}})]{intel-onednn}
{Intel oneAPI Deep Neural Network Library}, {\natexlab{b}}.
\newblock URL
  \url{https://www.intel.com/content/www/us/en/developer/tools/oneapi/onednn.html}.

\bibitem[int({\natexlab{c}})]{intel-oneapi}
{Intel oneAPI AI Analytics Toolkit}, {\natexlab{c}}.
\newblock URL
  \url{https://www.intel.com/content/www/us/en/developer/tools/oneapi/toolkits.html}.

\bibitem[Sergeev and Balso(2018)]{sergeev2018horovod}
Alexander Sergeev and Mike~Del Balso.
\newblock Horovod: fast and easy distributed deep learning in tensorflow, 2018.

\bibitem[Gabriel et~al.(2004)Gabriel, Fagg, Bosilca, Angskun, Dongarra,
  Squyres, Sahay, Kambadur, Barrett, Lumsdaine, Castain, Daniel, Graham, and
  Woodall]{open-mpi}
Edgar Gabriel, Graham~E. Fagg, George Bosilca, Thara Angskun, Jack~J. Dongarra,
  Jeffrey~M. Squyres, Vishal Sahay, Prabhanjan Kambadur, Brian Barrett, Andrew
  Lumsdaine, Ralph~H. Castain, David~J. Daniel, Richard~L. Graham, and
  Timothy~S. Woodall.
\newblock Open {MPI}: Goals, concept, and design of a next generation {MPI}
  implementation.
\newblock In \emph{Proceedings, 11th European PVM/MPI Users' Group Meeting},
  pages 97--104, Budapest, Hungary, September 2004.

\end{thebibliography}

\end{document}